\documentstyle[aps,preprint]{revtex}
\begin{document}

\title{Non-universal scaling and dynamical
feedback  in generalized models of financial markets }
\author{Dafang Zheng$^{1,2,}$\thanks
{e-mail: dafang.zheng@brunel.ac.uk, phdzheng@scut.edu.cn}, 
G. J. Rodgers$^{1}$, P. M. Hui$^{3}$ and R. D'Hulst$^{1}$}
\address{$^{1}$ Department of Mathematical Sciences, Brunel  University,\\
 Uxbridge, Middlesex UB8 3PH, UK}
\address{$^{2}$ Department of Applied Physics, South China University of
Technology,\\ Guangzhou 510641, P.R. China}
\address{$^{3}$ Department of Physics, The Chinese University of Hong
Kong,\\
Shatin, New Territories, Hong Kong}
\maketitle

\begin{abstract}

We study self-organized  models for 
information transmission and herd behavior 
in financial markets. Existing models are generalized to take into account the 
effect of size-dependent fragmentation and coagulation probabilities of 
groups of agents and to include a demand process.
Non-universal scaling with  a tunable exponent for the group 
size distribution is found in the resulting system. We also show that the fragmentation 
and coagulation probabilities of groups of agents have a strong influence on the average investment
rate of the system.

\vspace*{0.2 true in}


\end{abstract}

\newpage
\section{Introduction}

Recently, empirical studies have shown that  the price fluctuations measured by the returns in
financial markets have a heavy-tailed non-Gaussian distribution\cite{stanley,gopikrishnan}.
Understanding the microscopic mechanism of this market phenomena is a challenging 
problem that has recently attracted the interest of physicists\cite{bak,farmer,lux}. 
In a simple percolation model recently proposed by Cont and Bouchaud(CB)\cite{cont}, it is shown that
the {\em herd behavior} can lead to the desired fat tails for the distributions of price returns in 
financial markets. Herd means a collective phenomena or a crowd effect, and it assumes that
agents do not make decisions independently, but they are usually grouped into clusters of agents
sharing the same information and hence making a common decision\cite{topol,bannerjee}. 
In the CB model, a random
{\em communication} network of agents is constructed and the group size distribution of agents
is linked to the distribution of price returns. It is interesting that in such a simple model 
the group size distribution of agents has a power law decay with an exponential cut-off, 
which can be compared with the fat-tailed distribution of price returns in real markets\cite{cont}. 
To capture further such a herd behavior, several extensions of the CB model have been presented,
including the fundamentalist influence in the CB model considered by Chang and Stauffer\cite{chang}, 
the self-organized dynamical model of Egu\'{i}luz and Zimmermann(EZ)\cite{EZ,dHR1} and the democracy
and dictatorship models proposed by D'Hulst and Rodgers\cite{dHR2}.
The exponent
characterizing the group size distributions in all of these generalized versions was found to be
{\em robust} with the same value of $5/2$. Interestingly, however, it was found in a recent work that 
this scaling behavior can actually be changed\cite{zheng}. By introducing the size-dependent
fragmentation and coagulation rates in the generalized EZ model, the group
size distribution of agents was found to be model-dependent with a tunable exponent 
$\beta(\delta)=5/2-\delta$, where $\delta$ is the exponent characterizing the power law decay for
the probabilities of fragmentation and coagulation of groups of agents in the system\cite{zheng}.

In this work, we will consider how the demand process in the democracy and 
dictatorship models\cite{dHR2} is affected by the dynamical mechanism of the fragmentation and
coagulation of groups of agents in the system. As in Ref.\cite{zheng}, we also introduce a size-dependent
fragmentation rate $f_{i}=s_{i}^{-\delta}$ and a size dependent coagulation rate 
$c_{ij}=s_{i}^{-\delta} s_{j}^{-\delta}$ to the existing models. We study how the scaling 
behavior for the group size distribution of agents, together with its character 
distribution $Q(p)$ \cite{dHR2}, depends on the parameter $\delta$ in the resulting system. 
We will discuss both the regimes $0\leq\delta<1$\cite{zheng} and $\delta\geq1$.

The outline of this paper is as follows.
In section II, we define the models and present both  analytical and numerical results
for the group size distribution $n_{s}$. The character distribution $Q(p)$ of agents is also studied
numerically. Section III contains further discussion of the relationship between
the average investment rate $\bar{a}$ and the parameter $\delta$. 
A brief summary of our results and conclusions is also given in this section.

\section{The Models}

The models proposed in this paper are generalized versions of the democracy
and dictatorship models introduced by D'Hulst and Rodgers\cite{dHR2}, which aim to
describe not only the information transmission and herd behavior but also the
demand process in financial markets. We consider a system with a  total
of $N$ agents. Initially, each agent is given a microscopic parameter $p$ , which is a
random number chosen from a uniform distribution between $0$ and $1$. Agents are organized 
into groups  sharing the same value of $p$. Agents belonging to a group
have the same opinion and hence make the same decision at a given moment in time. 
At the beginning of the simulation, all the agents are isolated, i.e. each group has only
one agent. At each time step, two agents $i$ and $j$, with  associated number $p_{i}$
and $p_{j}$ respectively, are selected at random. A microscopic investment rate is defined
as $a_{ij}=|p_{i}-p_{j}|$ , i.e., with probability $a_{ij}$ agent $i$ and all other agents belonging 
to his group decide to make a transaction, i.e. to buy or to sell with equal probability. After the
transaction the group of agent ${i}$ is broken up into isolated agents with probability 
$s_{i}^{-\delta}$, with $\delta \geq 0$, or the group remains unchanged with probability
$1-s_{i}^{-\delta}$.
Once the group of agent $i$ is fragmented each agent in it
will be given a new microscopic parameter $p$ chosen randomly from a uniform distribution
between $p_{i}-R$ and $p_{i}+R$, with $0<R<0.5$. On the other hand, with probability $1-a_{ij}$
agent $i$ decides not to make a transaction and all the agents in the group of agent $i$ will 
communicate with the agents in group $j$, i.e., the two groups of agent $i$ and agent $j$
will combine to form a bigger group with probability $s_{i}^{-\delta} s_{j}^{-\delta}$, or remain
separate with probability $(1- s_{i}^{-\delta}s_{j}^{-\delta})$. Once the two groups of agent $i$ and agent $j$
are coagulated all agents in both the groups will be given a new common parameter $p_{ij}$. In the
original democracy and dictatorship models, $p_{ij}=(p_{i}+p_{j})/2$ and $p_{ij}=p_{i}$ 
for the  democracy  and  dictatorship models, respectively.

In the original democracy and dictatorship models, the group of agent $i$ is fragmented after a transaction, 
and conversely the two groups of agent $i$ and agent $j$ are coagulated if there is no transaction.
In the present work, instead, we introduce a size-dependent fragmentation rate $f_{i}=s_{i}^{-\delta}$ 
and a size-dependent
coagulation rate $c_{ij}=s_{i}^{-\delta} s_{j}^{-\delta}$ into the systems. It seems reasonable for us to introduce 
such a generalization in a sense that the size-dependent fragmentation and coagualtion rates can be used to mimic
some dynamical properties in real markets, such as the growth and bankruptcy of 
businesses\cite{nagel,ramsden}. That is to say, 
smaller businesses are easier to bankrupt, and, on another hand, larger businesses 
are more difficult to grow up\cite{dHR3}.
So the present models can also be extended to study the size distribution of businesses. 

For our present generalized versions we can also write down a master equation for the number $n_{s}(t)$ of groups
 with size $s$\cite{dHR1,dHR2,zheng}
\begin{equation}
N \frac{\partial n_{s}}{\partial t} =
-\bar{a} s^{1-\delta} n_{s} + \frac{(1-\bar{a})}{N} \sum_{r=1}^{s-1}
r^{1-\delta} n_{r} (s-r)^{1-\delta} n_{s-r}
- \frac{2(1-\bar{a})s^{1-\delta} n_{s}}{N} \sum_{r=1}^{\infty} r^{1-\delta}n_{r}
\label{eq:ns time evol}
\end{equation}
for $s > 1$. Here the constant investment rate $a$ 
which appeared in the original EZ's model\cite{EZ}, as well as in its generalized
version\cite{zheng}, has been replaced by an averaged investment rate $\bar{a}$ \cite{dHR2}.
Each term on the right hand side of Eq.(\ref{eq:ns time evol})
represents the consequence of a possible action of the agents.  The first
term describes the dissociation of a group of size $s$ after a
transaction is made.  The second term represents coagulation of two groups
to form a group of size $s$.  The third term represents the
combination of a group of size $s$ with another group.  For groups
of size unity ($s=1$), we have
\begin{equation}
N \frac{\partial n_{1}}{\partial t} =
\bar{a} \sum_{r=2}^{\infty} r^{2-\delta}n_{r}
- \frac{2 (1-\bar{a})n_{1}}{N} \sum_{r=1}^{\infty} r^{1-\delta}n_{r}.
\label{eq:n1 time evol}
\end{equation}
Here, the first term comes from the dissociation of groups
into isolated agents and the second term describes the combination of a
group of size unity with another group.  In the steady state,
$\frac{\partial n_{s}}{\partial t} = 0$, we have
\begin{equation}
s^{1-\delta} n_{s} = \frac{1-\bar{a}}{N\bar{a} + 2(1-\bar{a})\sum_{r=1}^{\infty} r^{1-\delta}n_{r}}
 \sum_{r=1}^{x-1} r^{1-\delta}
(s-r)^{1-\delta} n_{r} n_{s-r}
\label{eq:ns steady state}
\end{equation}
for $s>1$, and 
\begin{equation}
n_{1} = \frac{N\bar{a}}{2(1-\bar{a}) \sum_{r=1}^{\infty} r^{1-\delta} n_{r}}
 \sum_{r=2}^{\infty} r^{2-\delta} n_{s}.
\label{eq:n1 steady state}
\end{equation}

Using a generating function method\cite{wallace}, we can derive a self-consistent equation for the number
$n_{s}$ of groups of size $s$\cite{zheng}
\begin{equation} 
n_{s} \sim N \left[ \frac{ 4(1-\bar{a})[(1-\bar{a}) + \frac{N\bar{a}}{\sum_{r=1}^{\infty}
r^{1-\delta}n_{r}}]}
{[\frac{N\bar{a}}{\sum_{r=1}^{\infty} r^{1-\delta}n_{r}} + 2(1-\bar{a})]^{2}}
\right]^{s} 
s^{-(\frac{5}{2} - \delta)}.
\label{eq:resulting ns}
\end{equation} 
For the case of $\delta=0$, it reduces to the expression 
for the original democracy and dictatorship models\cite{dHR2}.
It indicates  that the power law scaling for the distribution of the group number $n_{s}$ is
non-universal, with a tunable exponent 
$\beta(\delta)=\frac{5}{2}-\delta$ for both the generalized democracy
and dictatorship versions. Notice that the power law decay is mediated by an exponential cut-off term in
Eq.[\ref{eq:resulting ns}].
For the democracy model, $\bar{a} \approx 0$, which leads to a unit value for the exponential cut off. While, for the
dictatorship model, $\bar{a} \approx 0.5$, and the exponential cut off will dominate for larger $s$\cite{dHR2}.

Figure 1 shows our simulation results for the normalized group size distribution  $n_{s}/n_{1}$ depending on
$s$, for various values of the parameter $\delta$. In this paper all simulation results are obtained in a time 
window of $t=10^{5} \sim 10^{6}$ for a system with $N=10^{4}$. Averages are taken over $32$ independent runs. 
For $\delta=0$, the results for the original democracy and dictatorship models are recovered, i.e., one obtains
a power law scaling with exponent $\beta \approx 5/2$ in both the models, but with a dominant exponential cut
off for larger $s$ in the dictatorship version. For $0\leq\delta\leq1$, we  also have a power law
scaling for the group size distribution $n_{s}$, but with a tunable exponent $\beta(\delta)=5/2-\delta$ in both
models. Thus the exponent $\beta$ of the power law scaling for the group size
distribution $n_{s}$ is no longer so {\em robust}\cite{EZ,dHR1,dHR2}, 
but becomes model dependent and 
hence non-universal. One finds that the simulation results for the group size distribution $n_{s}$
in the region of $0\leq\delta\leq1$ are in good agreement with the analytic expression given by 
Eq.[\ref{eq:resulting ns}].
For $\delta>1$, the scaling behavior of the the group size distribution $n_{s}$ disappears step by step 
as $\delta$ increases. This indicates that the mean-field analysis of the master equation is invalid and hence
the resulting expression in Eq.[\ref{eq:resulting ns}]  for the group size distribution $n_{s}$ is 
no longer available once the value
of  the parameter $\delta$ exceeds $1$. This break down of scaling  can be understood qualitatively as the 
coagulation rate is too small  to form large groups of agents for larger $\delta$.

An important feature found in Ref.\cite{dHR2} is that the distributions $Q(p)$ of the $p$'s
for the democracy  and  dictatorship models
in the steady state  are very different. $Q(p)dp$ is defined as the relative number
of the agents associated with a value of the microscopic parameter $p$ inside $(p, p+dp)$.
The results for $Q(p)$  are presented in Figure 2, for various values of the parameter $\delta$ and a given
range of $R=0.1$. For the democracy model, as shown in Figure 2a, the system is driven towards a coherent
state where the distribution $Q(p)$ is Gaussian-like. One finds that the amplitude of spread and hence the height 
of the peak of $Q(p)$ depends sensitively on $\delta$ . For $0\leq\delta\leq1$, the peak grows as $\delta$ increases.
For $\delta>1$, however, the peak drops as $\delta$ increases. This turns out to be that there is a continuous 
transition for the distribution $Q(p)$
occured at around $\delta=1$. The distribution $Q(p)$ for the dictatorship model is very different from
that obtained in the democracy model. As shown in Figure 2b, a spontaneous segregation into two equal sized
populations occurs, with almost one half of the agents associated with a value of $p$ near $0$ and the rest near $1$.
The distribution $Q(p)$ also depends on  $\delta$, i.e., $Q(p)$ becomes flat as $\delta$ increases. On the other hand,
there is no transition for the distribution $Q(p)$ as $\delta$ increases, contrary to what 
is seen in the democracy model.
\section{Discussion}
We have found so far that the size-dependent fragmentation and coagulation rates can have a strong influence not
only on the scaling behavior of the group distribution $n_{s}$, but also on the character distribution, $Q(p)$, of agents.
An agent's microscopic parameter $p$ is actually a characterization of the way the agent 
is perceived in the market,
rather than only the individuality of the agent, and hence the distribution $Q(p)$ determines the decision process in
the market\cite{dHR2}. 
The fact that the distribution $Q(p)$ depends on $\delta$ demonstrates that the mechanism of the 
fragmentation and coagulation of groups presented in this work has a strong effect of feedback on the 
market, which is also reflected in the dependence of the average investment rate $\bar{a}$ on the value 
of $\delta$ as shown in Figure 3. For the democracy model as shown in Figure 3a, 
the average investment rate $\bar{a}$ decreases continuously as $\delta$ increases for $0\leq\delta<1$. For
$\delta>1$, however, $\bar{a}$ increases with $\delta$. $\bar{a}$ has a minimum at around $\delta=1$. Hence
the dependence of the average investment rate $\bar{a}$ on the value of the parameter $\delta$ shown in 
Figure 3a is consistent with the dependence of $Q(p)$ on  $\delta$ in Figure 2a. For the dictatorship case as 
shown in Figure 3b, the average investment rate $\bar{a}$ decreases rapidly as $\delta$ increases from
$\delta=0$ to $\delta=2$ and becomes flat as  $\delta$ goes beyond $\delta\approx2$. Thus there
is no transition in the average investment rate $\bar{a}$. 
The results found in Figure 3b are also consistent with those in Figure 2b. 

We can explain  the above observations qualitatively. For the democracy case, it is the
competition between the fragmentation and the coagulation of groups that leads to the  
dependence of the character distribution $Q(p)$, and hence the average investment rate $\bar{a}$, on
the parameter $\delta$. That is to say, the smaller the
fragmentation rate is, the easier it is  to form big groups in the steady state. On the other hand, however,
it is difficult to form big groups in the steady state in the case of  a small coagulation rate. Hence for a
small value of $\delta$, the fragmentation rate is dominant, which results in a decreasing dependence
of $\bar{a}$ on $\delta$. When $\delta$ exceeds some value, which is about $1$ as observed   
in the simulation, the coagulation rate becomes dominant and hence leads to an increasing
dependence of $\bar{a}$ on $\delta$. For the dictatorship model, the fragmentation rate is no longer
dominant in case of small $\delta$ due to the large average {\em effective} fragmentation rate
$\bar{f}_{effect}=\bar{a}\bar{f}$. Hence the average investment rate $\bar{a}$ depends 
only sensitively on the coagulation rate, which leads to a decreasing dependence of $\bar{a}$ on
$\delta$ for all the values of $\delta\geq0$.

In summary, we have introduced size-dependent fragmentation and coagulation rates to the democracy and 
dictatorship models proposed recently by D'Hulst and Rodgers\cite{dHR2}. As in the generalized version of the
EZ model\cite{zheng}, non-universal scaling is found in the systems. The exponents characterizing the group size
distribution in both  democracy and dictatorship models turn out to be model dependent and hence
are no longer robust. In the original EZ model\cite{EZ}, as well as in its generalized version\cite{zheng},
an investment rate $a$ is given artificially and is fixed during the whole dynamical trade process. In our present
work, however, the microscopic investment rate $a_{ij}=|p_{i}-p_{j}|$ is inhomogeneous and  is a dynamical
parameter governing the demand process of the system. The most interesting feature of the present models is that 
the average investment rate $\bar{a}$ is strongly influenced by the 
dynamical mechanism of the fragmention and coagulation of groups of agents. Hence the new models 
seem to be better
at mimicing the information transmission and herd behavior, together with the demand process and the
dynamical feedback inherent in real markets.

\begin{center}
{\bf ACKNOWLEDGEMENTS}
\end{center}

DFZ, GJR and PMH acknowledge financial  
support from The China Scholarship Council, The Leverhulme Trust  and the 
Research Grants Council of the Hong Kong SAR Government under grant 
CUHK4241/01P, respectively.

\newpage \centerline{\bf FIGURE CAPTIONS}

\bigskip
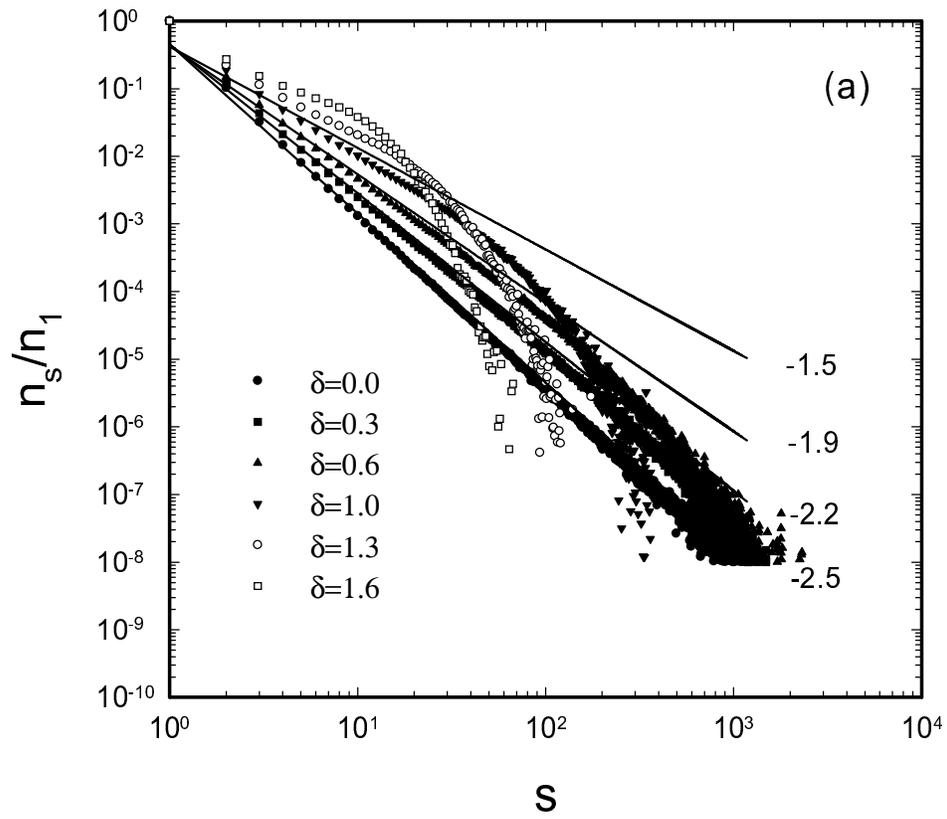
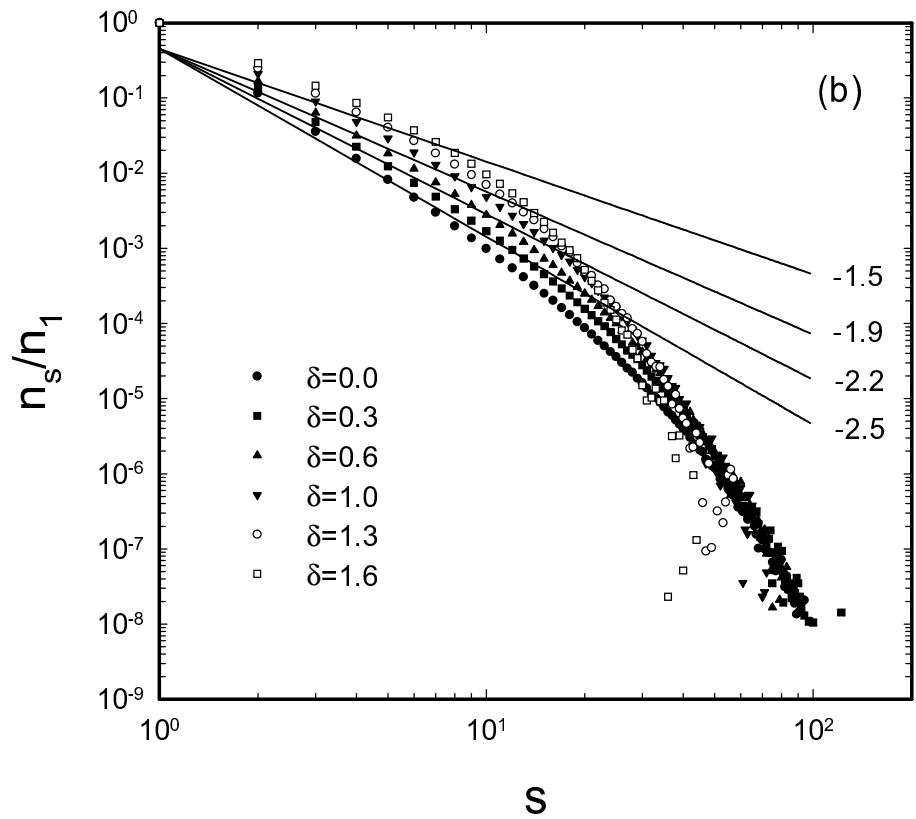
\noindent Figure 1: Normalized group size distribution $n_{s}/n_{1}$ as a function
of size $s$ on a log-log scale for (a) the generalized democracy model and 
(b) the generalized dictatorship model
for different values of
$\delta$ obtained by numerical simulations (symbols).  The values of
$\delta$ used in the calculations are: $\delta = 0, 0.30, 0.60, 1.0, 1.3, 1.6$.
The solid lines are a guide to the eye corresponding to exponents
$\beta = -2.5, -2.2, -1.9, -1.5$ respectively.

\bigskip
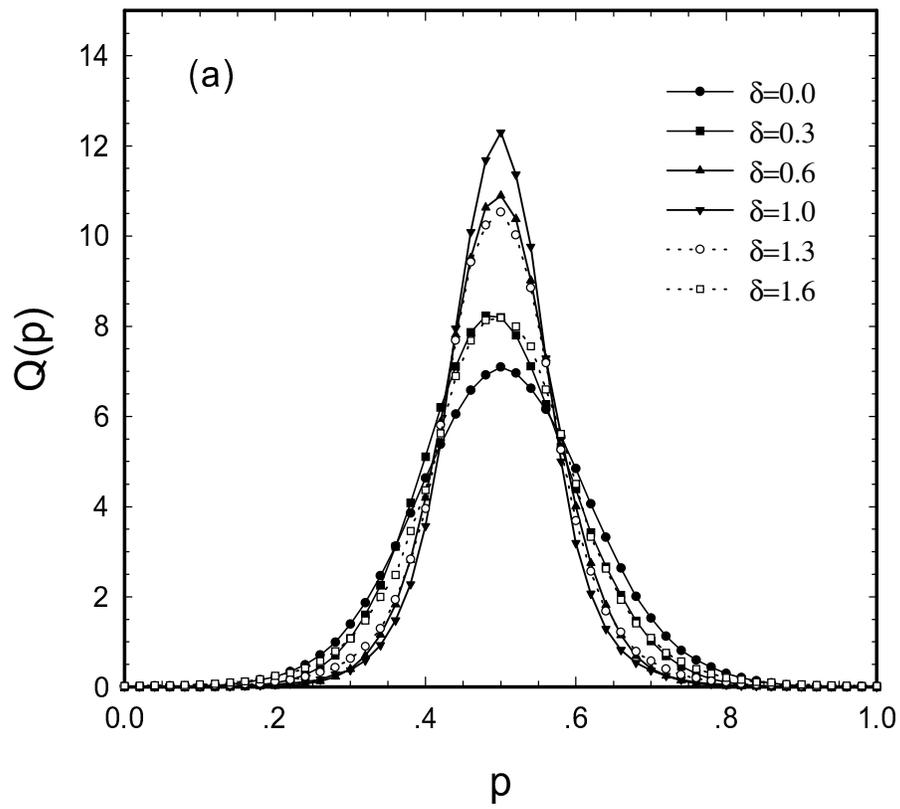
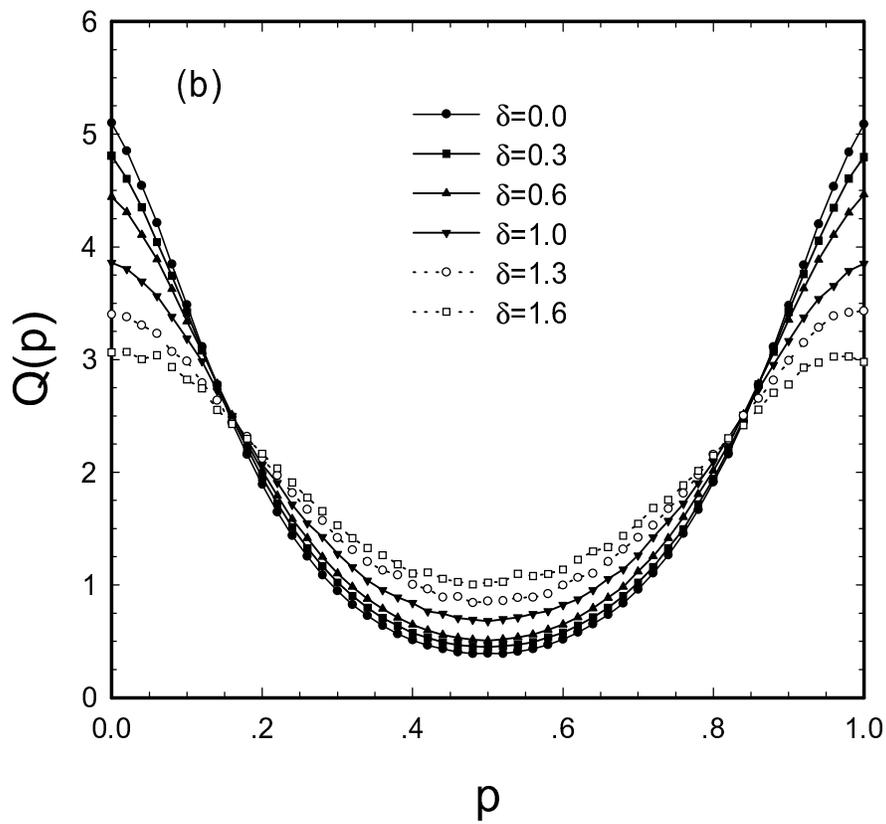
\noindent Figure 2: Probability distribution $Q(p)$ of the characters $p$ of the agents for
(a) the generalized democracy model and
(b) the generalized dictatorship model 
for different values of
$\delta$ obtained by numerical simulations.  The values of
$\delta$ used in the calculations are: $\delta = 0, 0.30, 0.60, 1.0, 1.3, 1.6$.

\bigskip
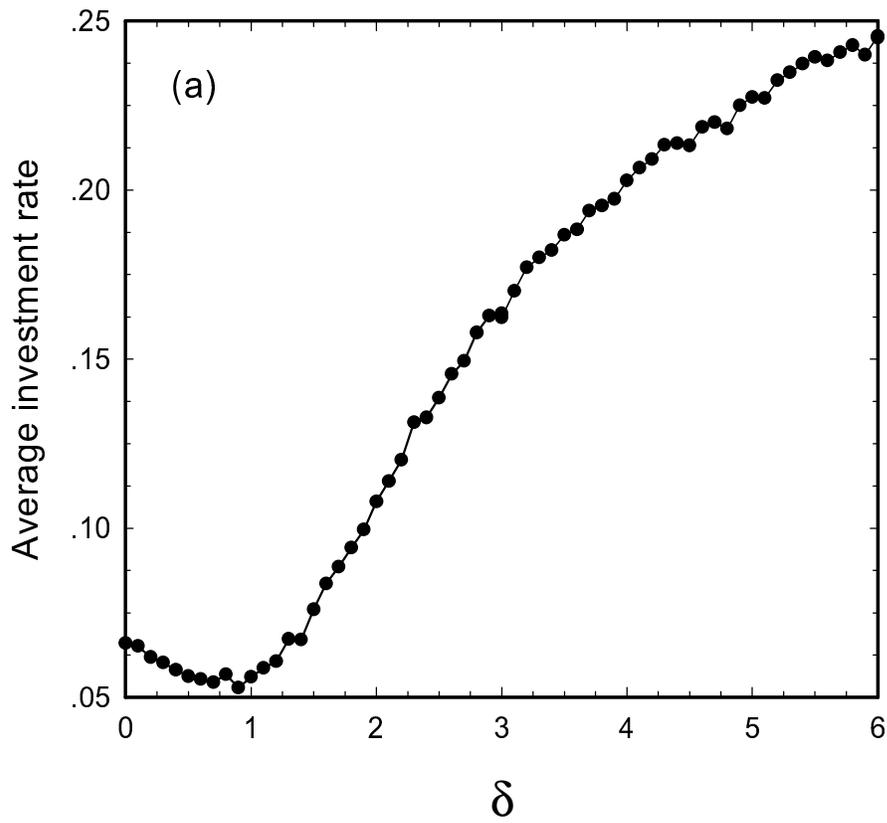
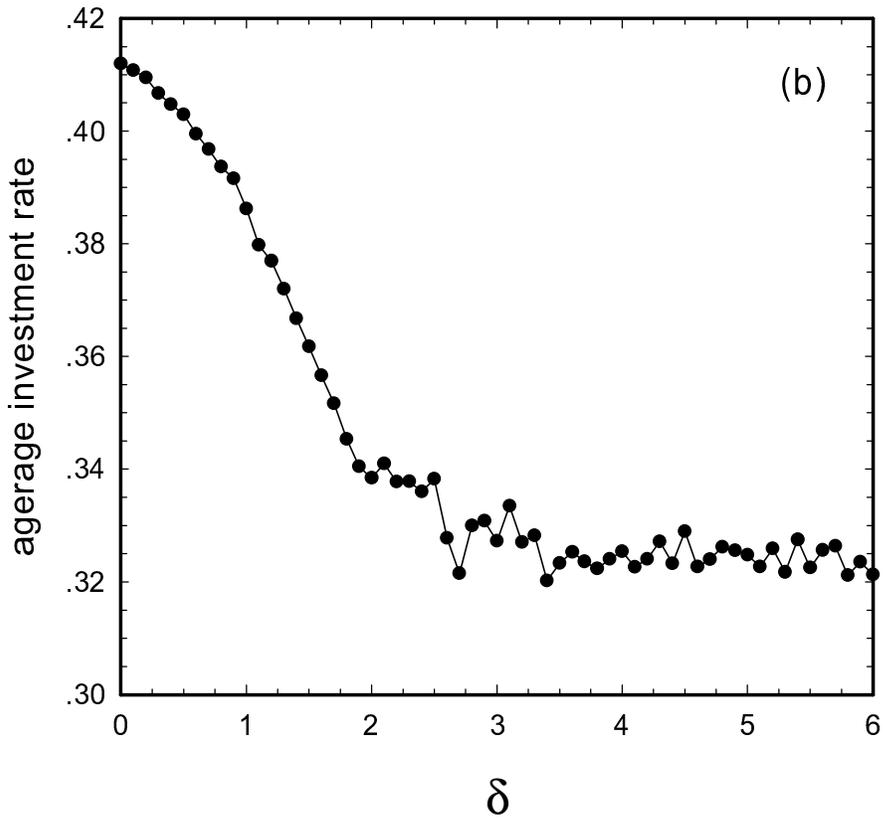
\noindent Figure 3: Average investment rate as a function of 
the parameter $\delta$ for
(a) the generalized democracy model and
(b) the generalized dictatorship model.

\end{document}